\documentclass[
    twocolumn,
	prd,
	amssymb,
	preprintnumbers,superscriptaddress,
	nofootinbib]{revtex4-1}

\pdfoutput=1
\usepackage{paralist}
\usepackage{graphicx}
\usepackage{enumitem}
\usepackage{latexsym}
\usepackage{amsfonts}
\usepackage{amssymb}
\usepackage{xcolor}
\usepackage[export]{adjustbox}
\usepackage{amsmath}
\usepackage[thinlines]{easytable}
\usepackage{slashed}
\usepackage{dcolumn}
\usepackage{verbatim}
\usepackage{float}
\usepackage{multirow}
\usepackage{xspace}
\usepackage[normalem]{ulem}
\usepackage[
pdfauthor={Djuna Croon}]{hyperref}
\usepackage{tabularx}
\usepackage{lettrine}
\usepackage{setspace}
\usepackage{cleveref}

\input Zallman.fd

\LettrineTextFont{\itshape}

\newcommand{\ra}{\rightarrow}

\newcommand{\acro}[1]{\textsc{\MakeLowercase{#1}}} 

\newcommand{\beq}{\begin{equation}}
\newcommand{\eeq}{\end{equation}}
\newcommand{\bea}{\begin{eqnarray}}
\newcommand{\eea}{\end{eqnarray}}
\newcommand{\nn}{\nonumber}

\newcommand{\refjnl}[1]{{\rm#1}}

\def\apj{\refjnl{ApJ}}                 

\usepackage{pdfbase}[2017/03/16]
\usepackage{xparse,ocgbase}
\usepackage{xcolor,calc}
\usepackage{tikzpagenodes,linegoal}
\usetikzlibrary{calc}
\usepackage{tcolorbox}

\ExplSyntaxOn
\let\tpPdfLink\pbs_pdflink:nn
\let\tpPdfAnnot\pbs_pdfannot:nnnn\let\tpPdfLastAnn\pbs_pdflastann:
\let\tpAppendToFields\pbs_appendtofields:n
\def\tpPdfXform{\pbs_pdfxform:nnnnn{1}{1}{}{}}
\let\tpPdfLastXform\pbs_pdflastxform:
\let\cListSet\clist_set:Nn\let\cListItem\clist_item:Nn
\ExplSyntaxOff

\usepackage{pdfbase}[2017/03/16]
\usepackage{xparse,ocgbase}
\usepackage{xcolor,calc}
\usepackage{tikzpagenodes,linegoal}
\usetikzlibrary{calc}
\usepackage{tcolorbox}

\ExplSyntaxOn
\let\tpPdfLink\pbs_pdflink:nn
\let\tpPdfAnnot\pbs_pdfannot:nnnn\let\tpPdfLastAnn\pbs_pdflastann:
\let\tpAppendToFields\pbs_appendtofields:n
\def\tpPdfXform{\pbs_pdfxform:nnnnn{1}{1}{}{}}
\let\tpPdfLastXform\pbs_pdflastxform:
\let\cListSet\clist_set:Nn\let\cListItem\clist_item:Nn
\ExplSyntaxOff

\makeatletter
\NewDocumentCommand{\tooltip}{%
  ssssO{\ifdefined\@linkcolor\@linkcolor\else blue\fi}mO{yellow!20}mO{0pt,0pt}%
}{{%
  \leavevmode%
  \IfBooleanT{#2}{%
    \ocgbase@new@ocg{tipOCG.\thetcnt}{%
      /Print<</PrintState/OFF>>/Export<</ExportState/OFF>>%
    }{false}%
    \xdef\tpTipOcg{\ocgbase@last@ocg}%
    \ocgbase@add@ocg@to@radiobtn@grp{tool@tips}{\ocgbase@last@ocg}%
  }%
  \tpPdfLink{%
    \IfBooleanTF{#4}{%
      /Subtype/Link/Border[0 0 0]/A <</S/SetOCGState/State [/Toggle \tpTipOcg]>>
    }{%
      /Subtype/Screen%
      /AA<<%
        \IfBooleanTF{#3}{%
          /E<</S/SetOCGState/State [/Toggle \tpTipOcg]>>%
        }{%
          \IfBooleanTF{#2}{%
            /E<</S/SetOCGState/State [/ON \tpTipOcg]>>%
            /X<</S/SetOCGState/State [/OFF \tpTipOcg]>>%
          }{
            \IfBooleanTF{#1}{%
              /E<</S/JavaScript/JS(%
                var fd=this.getField('tip.\thetcnt');%
                if(typeof(click\thetcnt)=='undefined'){%
                  var click\thetcnt=false;%
                  var fdor\thetcnt=fd.rect;var dragging\thetcnt=false;%
                }%
                if(fd.display==display.hidden){%
                  fd.delay=true;fd.display=display.visible;fd.delay=false;%
                }else{%
                  if(!click\thetcnt&&!dragging\thetcnt){fd.display=display.hidden;}%
                  if(!dragging\thetcnt){click\thetcnt=false;}%
                }%
                this.dirty=false;%
              )>>%
            }{%
              /E<</S/JavaScript/JS(%
                var fd=this.getField('tip.\thetcnt');%
                if(typeof(click\thetcnt)=='undefined'){%
                  var click\thetcnt=false;%
                  var fdor\thetcnt=fd.rect;var dragging\thetcnt=false;%
                }%
                if(fd.display==display.hidden){%
                  fd.delay=true;fd.display=display.visible;fd.delay=false;%
                }%
               this.dirty=false;%
              )>>%
              /X<</S/JavaScript/JS(%
                if(!click\thetcnt&&!dragging\thetcnt){fd.display=display.hidden;}%
                if(!dragging\thetcnt){click\thetcnt=false;}%
                this.dirty=false;%
              )>>%
            }%
            /U<</S/JavaScript/JS(click\thetcnt=true;this.dirty=false;)>>%
            /PC<</S/JavaScript/JS (%
              var fd=this.getField('tip.\thetcnt');%
              try{fd.rect=fdor\thetcnt;}catch(e){}%
              fd.display=display.hidden;this.dirty=false;%
            )>>%
            /PO<</S/JavaScript/JS(this.dirty=false;)>>%
          }%
        }%
      >>%
    }%
  }{{\color{#5}#6}}%
  \sbox\tiptext{%
    \IfBooleanT{#2}{%
      \ocgbase@oc@bdc{\tpTipOcg}\ocgbase@open@stack@push{\tpTipOcg}}%
    \tcbox[colframe=black,colback=#7,size=fbox,arc=1ex,sharp corners=southwest]{#8}%
    \IfBooleanT{#2}{\ocgbase@oc@emc\ocgbase@open@stack@pop\tpNull}%
  }%
  \cListSet\tpOffsets{#9}%
  \edef\twd{\the\wd\tiptext}%
  \edef\tht{\the\ht\tiptext}%
  \edef\tdp{\the\dp\tiptext}%
  \tipshift=0pt%
  \IfBooleanTF{#2}{%
    \setlength\whatsleft{\linegoal}%
  }{%
    \measureremainder{\whatsleft}%
  }%
  \ifdim\whatsleft<\dimexpr\twd+\cListItem\tpOffsets{1}\relax%
    \setlength\tipshift{\whatsleft-\twd-\cListItem\tpOffsets{1}}\fi%
  \IfBooleanF{#2}{\tpPdfXform{\tiptext}}%
  \raisebox{\heightof{#6}+\tdp+\cListItem\tpOffsets{2}}[0pt][0pt]{%
    \makebox[0pt][l]{\hspace{\dimexpr\tipshift+\cListItem\tpOffsets{1}\relax}%
    \IfBooleanTF{#2}{\usebox{\tiptext}}{%
      \tpPdfAnnot{\twd}{\tht}{\tdp}{%
        /Subtype/Widget/FT/Btn/T (tip.\thetcnt)%
        /AP<</N \tpPdfLastXform>>%
        /MK<</TP 1/I \tpPdfLastXform/IF<</S/A/FB true/A [0.0 0.0]>>>>%
        /Ff 65536/F 3%
        /AA <<%
          /U <<%
            /S/JavaScript/JS(%
              var fd=event.target;%
              var mX=this.mouseX;var mY=this.mouseY;%
              var drag=function(){%
                var nX=this.mouseX;var nY=this.mouseY;%
                var dX=nX-mX;var dY=nY-mY;%
                var fdr=fd.rect;%
                fdr[0]+=dX;fdr[1]+=dY;fdr[2]+=dX;fdr[3]+=dY;%
                fd.rect=fdr;mX=nX;mY=nY;%
              };%
              if(!dragging\thetcnt){%
                dragging\thetcnt=true;Int=app.setInterval("drag()",1);%
              }%
              else{app.clearInterval(Int);dragging\thetcnt=false;}%
              this.dirty=false;%
            )%
          >>%
        >>%
      }%
      \tpAppendToFields{\tpPdfLastAnn}%
    }%
  }}%
  \stepcounter{tcnt}%
}}
\makeatother
\newsavebox\tiptext\newcounter{tcnt}
\newlength{\whatsleft}\newlength{\tipshift}
\newcommand{\measureremainder}[1]{%
  \begin{tikzpicture}[overlay,remember picture]
    \path let \p0 = (0,0), \p1 = (current page.east) in
      [/utils/exec={\pgfmathsetlength#1{\x1-\x0}\global#1=#1}];
  \end{tikzpicture}%
}

\newcommand{\DLensSource}{D_{\rm LS}}
\newcommand{\DLens}{D_{\rm L}}
\newcommand{\DSource}{D_{\rm S}}
\newcommand{\uT}{u_{1.34}}
\newcommand{\thetaE}{\theta_{\rm E}}
\newcommand{\rE}{r_{\rm E}}
\newcommand{\tE}{t_{\rm E}}
\newcommand{\vE}{v_{\rm E}}

\newcommand{\der}{{\rm d}}

\allowdisplaybreaks

\begin{document}

\title{Microlensing signatures of extended dark objects using machine learning}

\author{Miguel Crispim Romao}
\email{miguel.romao@durham.ac.uk}
\affiliation{Institute for Particle Physics Phenomenology, Department of Physics, Durham University, Durham DH1 3LE, U.K.}
\author{Djuna Croon} \email{djuna.l.croon@durham.ac.uk}
\affiliation{Institute for Particle Physics Phenomenology, Department of Physics, Durham University, Durham DH1 3LE, U.K.}

\date{\today}

\begin{abstract}
	This paper presents a machine learning-based method for the detection of
	the unique gravitational microlensing signatures of extended dark objects, such as boson stars, axion miniclusters and subhalos.
	We adapt MicroLIA, a machine learning-based package tailored to handle the challenges posed by low-cadence data in microlensing surveys.
	Using realistic observational timestamps, our models are trained on simulated light curves to distinguish between microlensing by point-like and extended lenses, as well as from other object classes which give a variable magnitude.
	We focus on boson stars and NFW-subhalos and show that the former, which are examples of objects with a relatively flat mass distribution, can be confidently identified for $0.8 \lesssim r/r_E\lesssim 3$.
	Intriguingly, we also find that more sharply peaked structures, such as NFW-subhalos, can be distinctly recognized from point-lenses under regular observation cadence.
	Our findings significantly advance the potential of microlensing data in uncovering the elusive nature of extended dark objects.
	The code and dataset used are also provided.
\end{abstract}

\preprint{IPPP/24/02}

\maketitle

\section{Introduction}
Macroscopic dark matter candidates, with masses ranging from large asteroids ($\sim 10^{-15} \rm M_\odot$) to stars ($\sim \rm M_\odot$), offer compelling alternatives to the traditional particle-based theories. These celestial objects, potentially formed in the early universe, are primarily detectable via their gravitational effects: gravitational lensing
(e.g. \cite{Alcock_1998,Niikura:2017zjd,Niikura:2019kqi,Green:2020jor}) and gravitational waves (e.g. \cite{Bird:2016dcv,Ali-Haimoud:2017rtz,Kavanagh:2018ggo,Bertone:2019irm,LIGOScientific:2019kan,Chen:2019irf,Franciolini:2021nvv,Croon:2022tmr}).
Indeed, gravitational microlensing is one of the most important ways of probing compact objects such as  ``machos" or primordial black holes (PBHs), a dark matter (DM) candidate consisting of compact objects formed in the early Universe. Through surveys of a range of sources, microlensing of such point-like lenses has been used to constrain the fraction of DM such objects can comprise in a wide range of masses (see e.g. \cite{Green:2020jor}).

It has also been proposed that dark matter can instead be comprised of extended objects, such as boson stars (e.g.~\cite{Ruffini:1969qy,Gleiser:1988rq,BSReview,Croon:2018ybs}), axion miniclusters~\cite{Fairbairn:2017dmf}, and subhalos~\cite{ErickcekKris,Barenboim:2013gya,Fan2014,CoDecay,VectorDMInflation}.
Like compact objects, such objects can also bend the light of distant stars. Whether this effect can be probed by a microlensing survey depends on the comparison between the object radius and the Einstein radius -- the characteristic length scale which is a function of the mass of the dark matter lens and the distance to the light source.
The effectiveness of (micro-)lensing then depends on the size of the object compared to the Einstein radius: dilute dark objects, which are transparent to light, are ineffective lenses. Using conservative assumptions about the number of events observed,  Refs.\cite{Croon:2020wpr,Croon:2020ouk} derived modified constraints on extended dark matter objects.

\begin{figure}
	\includegraphics[width=.475\textwidth]{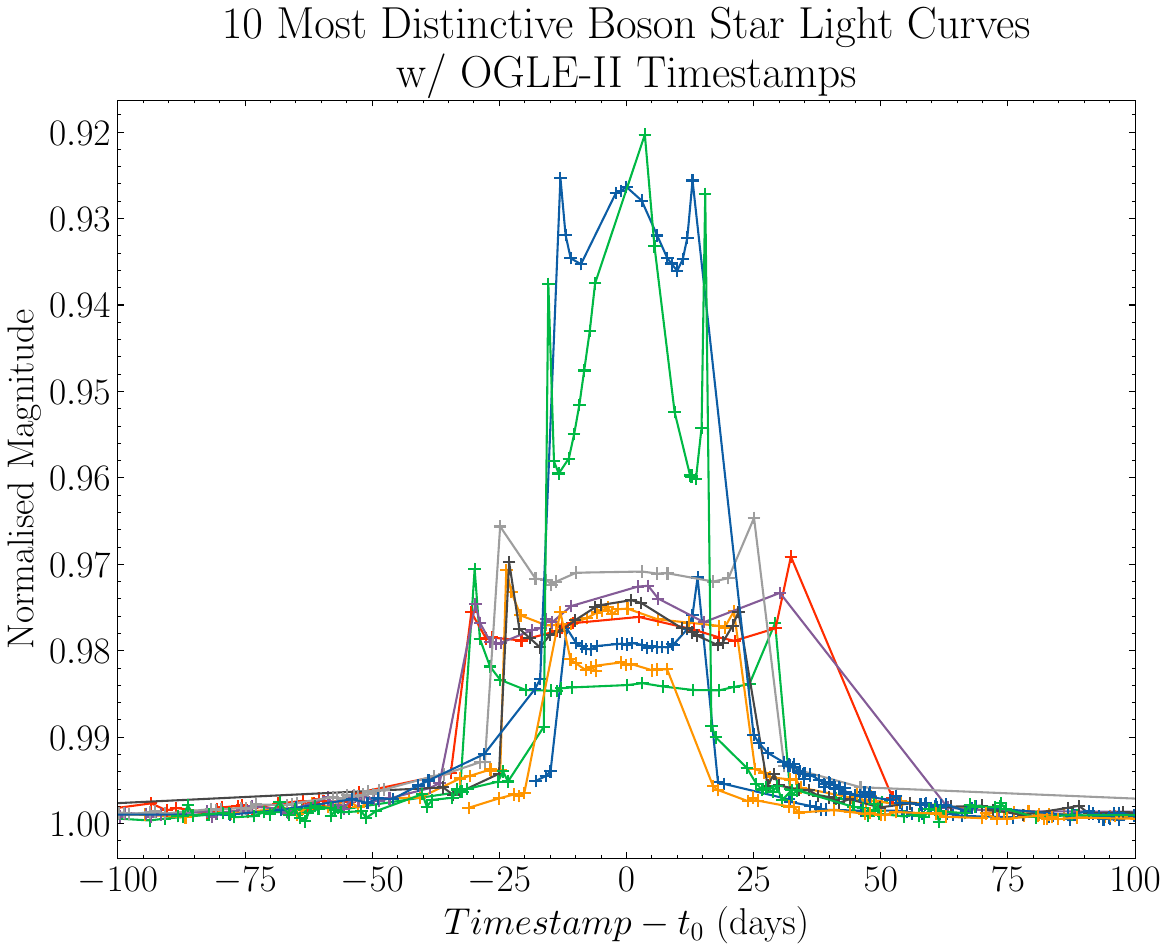}
	\caption{The 10 most distinctive Boson Star light curves, using the dataset generated with OGLE-II timestamps..
	}
	\label{Fig:10_most_distinctive_BS_lc_OGLEII}
\end{figure}

Interestingly, structures with radii close to the Einstein radius may give distinct microlensing signatures.
How the mass is spread within these structures affects the lensing effect.
This was demonstrated explicitly for microlensing of various dark matter structures in \cite{Croon:2020ouk,Bai:2020jfm,Croon:2020wpr}.
In objects with a flatter density profile, such as boson stars, the microlensing magnification time series can deviate significantly from that expected from a point-like lens such as a PBH, for example featuring caustics crossings as can be seen in Fig.~\ref{Fig:10_most_distinctive_BS_lc_OGLEII}.
These distinguishable features of extended dark lenses can in principle be used to make a positive discovery.

In this work we take the first step towards a positive discovery of a microlensing signature by an extended dark object.
To this end we develop an analysis pipeline to search for signatures of extended dark matter objects in time series data. Specifically, we train a histogram-based gradient boosted classifier to identify such features, and describe how these observations can be used to search for dark objects in a range of experiments.

This paper is organised as follows. We first review microlensing by extended lenses and estimate the sensitivity of the OGLE survey to the modified light curves of boson stars. We then describe the generation of our dataset and methodology, followed by an analysis of our results and a discussion.

\section{Microlensing signatures of extended lenses}
Let us first review some basics of gravitational microlensing, largely     following the treatment in Ref.~\cite{Narayan:1996ba}.
We will need to define some important parameters in the lens setup; a depiction of the geometry can be found in \cite{Croon:2020wpr}.
The observer-lens, lens-source, and observer-source distances are denoted $\DLens$, $\DSource$, and $\DLensSource$ respectively.
From the perspective of the observer, the lens center subtends angles of $\beta$ and $\theta_i$ with the source and images of the source respectively.
As $\DLens$, $\DSource$, and $\DLensSource$ are much larger than all other scales in the problem, lensing calculations can be simplified using the small angle approximation. In this approximation, the deflections $\alpha = 4 GM/(c^2 \xi)$ only occur when starlight encounters the ``lens plane'' perpendicular to the observer-source axis.

Assuming that the lens is spherically symmetric with density distribution $\rho(r)$ (and total mass $M=4\pi\int_0^\infty dr r^2\rho(r)$), the lensing equation may be written
\beq
\beta=\theta-\frac{\thetaE^2}{\theta}\frac{M(\theta)}{M}~,
\label{eq:lens}
\eeq
with the surface mass density projected onto the lens plane as
\bea
\nn M(\theta)&=&2\pi \DLens^2\int_0^\theta d\theta^\prime \theta^\prime \Sigma(\theta^\prime)~,\\
\Sigma(\theta)&=&\int_{-\infty}^\infty dz\,\rho\left(\sqrt{\DLens^2\theta^2+z^2}\right)~,
\label{eq:Mthetasigmatheta}
\eea
and where~\cite{Einstein:1956zz}
\begin{equation}
	\thetaE \equiv\sqrt{\frac{4GM}{c^2}\frac{\DLensSource}{\DLens\DSource}}~,
	\label{eq:thetabar}
\end{equation}
is the point-like Einstein angle, the value of $\theta$ for a point-like lens ($M(\theta) \ra M$) at zero impact parameter ($\beta = 0$).
This in turn defines the point-like Einstein radius $\rE \equiv \DLens \thetaE$ on the lens plane.

As the only way in which the total lens mass $M$ and the distances $\DLens, \DLensSource, \DSource$ enter the problem is through their contributions to the Einstein radius $\rE$, it is convenient to express all angles in units of $\thetaE$ and all distances $ \rE$.
Thus, we define $u \equiv \beta/\thetaE= \DLens \beta/\rE$, and $\tau \equiv \theta/\thetaE = \DLens \theta/\rE$ {(note that the latter was named $t$ in \cite{Croon:2020ouk,Croon:2020wpr})}, which allows us to rewrite ~\eqref{eq:lens}  as
\begin{equation}
	u=\tau-\frac{m(\tau)}{\tau},
	\label{eq:lens2}
\end{equation}
where we have also defined $m(\tau) \equiv M( \thetaE \tau)/M$ which describes the distribution of the lens mass projected onto the lens plane,
\beq
m(\tau) = \frac{\int_0^{\tau} d\sigma \sigma \int^\infty_0 d\lambda\, \rho(\rE\sqrt{\sigma^2+\lambda^2})}{\int_0^\infty d\gamma \gamma^2 \rho(\rE\gamma)}~,
\label{eq:mtmaster}
\eeq
where $\rho$ is the density distribution of the lens.

The lensing equation ~\eqref{eq:lens2} can be used to find the position(s) of the images $\theta_i$ given a position of the lens $\beta$.
As the images subtend different solid angles than the unlensed source, microlensing alters the observed flux.
The magnification is the ratio of the angular extent of an image to that of the source:
\bea
\begin{split}
	\mu = & \sum_i \left|\frac{\theta_i}{\beta}\frac{d\theta_i}{d\beta}\right|
	= \sum_i \left|\frac{\tau_i}{u}\frac{d\tau_i}{du}\right|                                                                                                 \\
	=     & \sum_i \left|1-\frac{m(\tau_i)}{\tau_i^2}\right|^{-1}\left|1+\frac{m(\tau_i)}{\tau_i^2}-\frac{1}{\tau_i}\frac{dm(\tau_i)}{d\tau_i}\right|^{-1}~.
\end{split}
\label{eq:mag}
\eea
The light curve as a function of time $t$ for a lens with velocity $v$ and minimum impact parameter $b_{\rm min}$ can now be found through $\beta D_L = \sqrt{b^2_{\rm min} + v^2t^2}$.
Whether a microlensing event is observable depends on the minimum detectable magnification for a given telescope, as well as the range of cadences at the microlensing survey, which sets the transit timescales to which it is sensitive. Typically, a transit is counted as a lensing event if $\mu >1.34$, which occurs for a point-like lens for impact parameter (in units of the Einstein radius) $u \equiv b/\rE = \beta/\thetaE = 1$

In \cite{Croon:2020wpr,Croon:2020ouk}
the microlensing efficiency of an extended lens compared to that of a point-like lens was defined as the maximum impact parameter $\uT$ for which a threshold magnification is produced: $\mu_{\rm tot}(u\le \uT) \ge 1.34~.$ For a point-like lens, $\uT =1$ by definition, and the naive expectation for extended lenses would be that $\uT \leq 1$. Remarkably, this is not necessarily the case: in particular for lenses with a reasonably flat density profile, $\uT$ can be larger than one.
Given $\uT$, the microlensing differential event rate for a single source with respect to the typical event timescale $t_E$ and $x = \DLens/\DSource$, observed by a particular experiment, can be calculated as
\beq
\frac{d^2\Gamma}{dx d\tE} = \varepsilon(\tE) \frac{2\DSource}{v_0^2 M} f_{\rm DM} \rho_{\rm DM}(x) \vE^4(x) e^{-\vE^2(x)/v^2_0}~,
\label{eq:dGammadxdt}
\eeq
where $\varepsilon(\tE)$ is the efficiency of telescopic detection, $\vE(x) \equiv 2 \uT(x)\rE(x)/\tE$, $v_0 = 220$~km/s is the dark matter circular speed in the galaxy,   and $\rho_{\rm DM} (x)$ is the DM density projected onto the line of sight, for example following an isothermal profile in the Milky Way galaxy,
\begin{equation}
	\begin{split}
		\nn \rho_{\rm DM} (r) & = \frac{\rho_{\rm s}}{1+(r/r_{\rm s})^2}~,                                                 \\
		r                     & \equiv \sqrt{R^2_{\rm Sol} - 2 x R_{\rm Sol} \DSource \cos \ell \cos b + x^2 \DSource^2}~,
	\end{split}
\end{equation}
with $R_{\rm Sol}$ = 8.5~kpc, $\rho_{\rm s} = 1.39$~GeV/cm$^3$, and $r_{\rm s} = 4.38$~kpc~\cite{PPPPCookbook};
$\ell$ and $b$ are the longitude and latitude of the source in galactic co-ordinates.
The total number of events is then given by
\beq
N_{\rm events} = N_\star T_{\rm obs} \int^1_0 dx \int^{t_{\rm E, max}}_{t_{\rm E, min}} d\tE \frac{d^2\Gamma}{dx d\tE}~,
\label{eq:Nevents}
\eeq
where
$N_\star$ is the number of observed sources, $T_{\rm obs}$ is the total observation time,
and $t_{\rm E, min}$ ($t_{\rm E, max}$) is the minimum (maximum) timescale of an event.

\begin{figure}
	\centering
	\includegraphics[width=.48\textwidth]{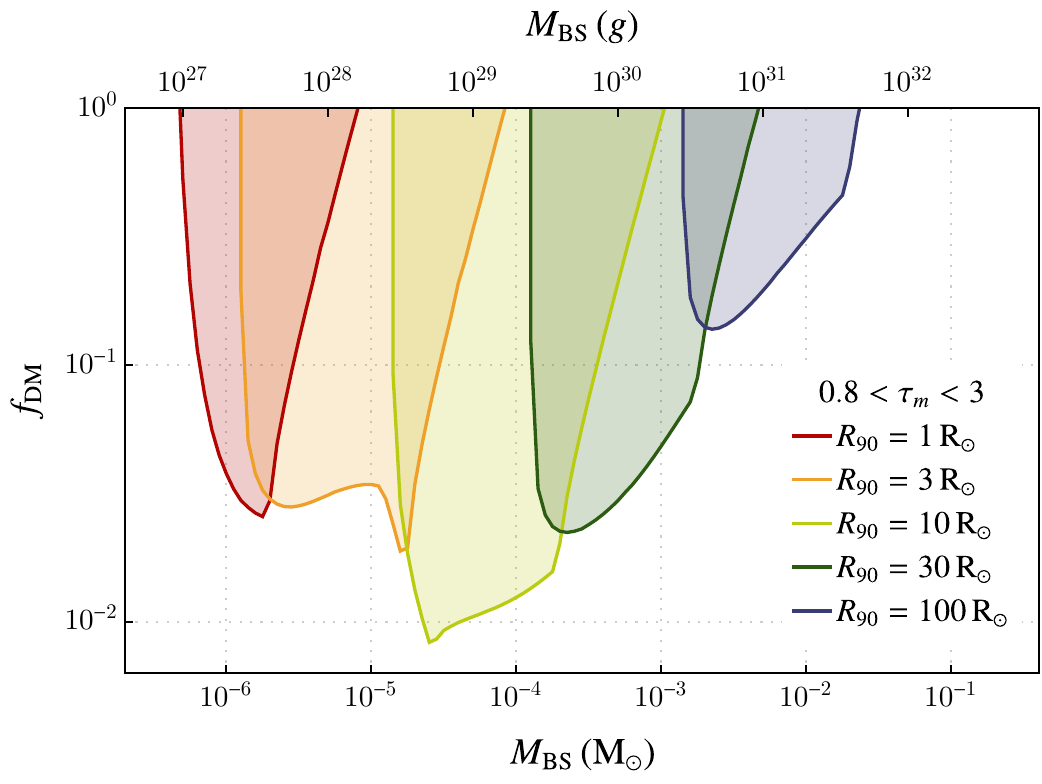}
	\caption{Sensitivity of the OGLE survey to a modified lensing light curve of a boson star. Here we anticipate that a boson star can be distinguished from a point-like lens for $0.8<\tau_m<3$.
	}
	\label{Fig:sensitivity}
\end{figure}

Comparing $N_{\rm events}$ with the number of observed events in microlensing surveys, \cite{Croon:2020wpr,Croon:2020ouk} set constraints on extended lenses. However, it is important to point out that the surveys identify microlensing events based on a comparison with the point-like magnification light curve. For extended lenses, in particular lenses with $ \tau_m \equiv r_{\rm lens}/r_E \sim 1 $, this is not a good approximation.
In Fig.~\ref{Fig:sensitivity} we estimate the range of lens masses and radial sizes for which we expect significant deviations from the point-like light curve, for the example of a boson star observed in the OGLE-IV survey (5 year dataset) ~\cite{Niikura:2019kqi}, anticipating that the non-compactness of the lens can be resolved for $0.8 < \tau_m < 3$ (which we will verify in the next section). Here we follow the mass profile of boson stars $m(t)$ outlined in the appendix of \cite{Croon:2020wpr}.
As in this work, the sensitivity is computed using the Poissonian $90\%$ confidence limit based on a signal comprised of dark matter and astrophysical foregrounds (see \cite{Niikura:2019kqi}).
We note that as the OGLE collaboration did not search for the particular microlensing light curves predicted by boson stars, this is an estimate only.
Comparing Fig.~\ref{Fig:sensitivity} with the sensitivity of the OGLE-IV survey to boson stars in \cite{Croon:2020wpr}, we note that for a given lens size, the lighter masses lead to distinguishable features in the light curve, as can be expected from the dependence of the Einstein radius on $M$.

Because the Einstein radius varies along the line of sight to the source, for a lens of a particular size $r_{\rm lens}$, a range of $\tau_m(x)$ is relevant, where $x=\DLens/\DSource$.
One might wonder what the distribution in $\tau_m$ is. This is given by $ \der N/ \der \tau_m = (\der N / \der x ) (\der x / \der \tau_m) = f_{\rm DM} \rho_{\rm DM} (x) \; \der x / \der \tau_m$. For the OGLE sources, we find that this distribution peaks for $ \tau_m (x=0.5)$, and rapidly falls like $\tau_m^{-3}$ away from it.
Thus, we expect that given a particular lens mass, it is in practise possible to interpret a measurement of $\tau_m$ directly in terms of a lens size.

\section{Dataset Generation and Methodology}
In this work we adapt MicroLIA, a tool developed and detailed in \cite{godines2019machine}.
The classifier developed in this paper is a machine learning model utilising the Random Forest algorithm, specifically designed for the detection of microlensing events in astronomical surveys. It is tailored to handle the challenges posed by low-cadence data, which typically suffer from irregular signal sampling and thus lower signal-to-noise ratios, making microlensing event detection more difficult. MicroLIA distinguishes between microlensing and other variable star events using 148 features derived from the light curve and its derivative time series.
The classifier categorises events into classes like microlensing, eclipsing binaries, and regular variable stars, focussing on accurate identification of microlensing amidst these.

In this work, we extend the scope of MicroLIA by including {extended dark matter objects which act like} non-point-like microlensing {lenses,}
such as Boson Stars (BS, focusing here on non-relativistic condensates of massive bosons without self-interactions) and Navarro-Frenk-White subhalos (NFW). For this, we generated microlensing light curves for these extended objects using their mass profile, $m(\tau)$, which have previously been calculated in \cite{Croon:2020wpr}. From these mass profiles, we fit an interpolating function, and for each value of the impact parameter, $u$, we solve the microlensing equation~\ref{eq:lens2} to obtain the total magnification to the light curve,~\ref{eq:mag}.

The observed impact parameters are closely related to the cadence of a survey. For a lens with
	{a characteristic timescale defined by the crossing of the Einstein radius} $t_E{\equiv R_E/v}$
and a minimum impact parameter $u_0$ producing a magnification peak occurring at $t_0$, the values of the observed impact parameters are
\begin{equation}
	u = \sqrt{u_0^2 + \left(\frac{t-t_0}{t_E}\right)^2} \ ,
\end{equation}
where $t$ is the survey time.
Therefore, the cadence at which the survey collects observational data will have a significant impact in the observed light curve. For this study, we considered two possible cases for the data collection timestamps. In the first case we used OGLE-II timestamps (but not the light curves nor their errors), which are input into MicroLIA and randomly sampled when simulating individual light curves.
	{We note that the regularity of the cadence in OGLE-II is not significantly different from other iterations of the survey. }
In the second case we considered a \emph{perfectly regular daily cadence}, where timestamps are all equally spaced by the same interval. As we will see below, this case demonstrates some identification opportunities that are obscured by irregular cadences, such as the case with OGLE-II timestamps.
We will refer to these to cases as OGLE-II Timestamps  and Regular Daily Cadence, respectively.

Both datasets were generated by simulating 100 000 light curves for each of the six classes: Cataclysmic Variables (CV), RR Lyrae and Cepheid Variables (VARIABLE), Mira long-period variables (LPV), Point-like Microlensing (ML), Boson Stars (BS), and NFW Subhalos (NFW). The CV, VARIABLE, LPV, and ML light curves were generated using MicroLIA's simulation, and BS and NFW were generated using our own simulation. We applied the same selection criteria for the three microlensing source events (ML, BS, NFW). The criteria are the same as MicroLIA's, but we further demanded that the observed magnification be at least $1.34$, {a common criterion imposed by microlensing surveys}. The light curves were simulated with a minimum magnitude of 15, a maximum magnitude of 20, and with Gaussian noise.

The extended microlensing sources, BS and NFW, have mass profiles which depend on the parameter $\tau_m$, which follows some distribution depending on the prevalence of such objects. For this study, we sampled $\tau_m \sim \mathcal{U}(0.5,5)$ logarithmically, since, as discussed in the previous section, this distribution is strongly peaked at the $\tau_m$ reached at $x=0.5$, and the largest deviations in the light curves are expected for $\tau_m \sim 1$.
Finally, we sampled the minimal impact parameter, $u_0$, differently for each of the three microlensing sources
\begin{equation}
	u_0 \sim \begin{cases}
		\mathcal{U}(0,1)   & \text{for ML}      \\
		\mathcal{U}(0,1.5) & \text{for BS}      \\
		\mathcal{U}(0,1.1) & \text{for NFW} \ ,
	\end{cases}
\end{equation}
where we emphasise that these are used during the simulation step, i.e. before the selection criteria, including $\mu \geq 1.34$, are imposed.

In total, we generated 600 000 curves, of which only a few $\mathcal{O}(0.1\%)$ with missing values for some of the features
	{were}
dropped. The dataset was then split into train, validation, and test subsets with proportions 0.5:0.25:0.25, respectively corresponding to around 300 000, 150 000, and 150 000 light curves, with the six classes being equally represented in each of the sets. For each light curve, we computed 74 features using the light curve time series, in addition to the same 74 using the derivative of the light curve time series, to a total of 148 features.
The computed features relate to statistics of the time series of the light curve, as well as other time series quantities. See~\cite{godines2019machine} and MicroLIA's API reference for the full description and reference of each feature.\footnote{\url{https://microlia.readthedocs.io/en/latest/autoapi/MicroLIA/features/}} The training set was used to conduct exploratory data analysis and to train machine learning models. The validation set was used for model selection and comparison, and to produce preliminary analysis plots and statistics. The test set was used to produce the final analysis plots and statistics, presented in the next section.~The full dataset can be downloaded here~\cite{crispim_romao_2024_10566869}.

For the multiclassification task, we trained a histogram-based gradient boosted classifier, which is a member of the broader machine learning algorithm family of gradient boosting machines (GBM). Like Random Forests (RF), implemented in MicroLIA, GBM are ensemble models that leverage the power of multiple weaker estimators, usually small trees, to produce a strong estimator. In RF the trees are independent of each other and the final prediction is obtained by the ensemble average over all the predictions from the trees in the forest. In GBM the trees are not independent but sequentially trained as to improve on the previous iteration. Schematically, consider $F(X)$ to be the output of the GBM as a function of the data, $X$. In the first iteration, we want to train a simpler estimator, $F_1(X)$, to match a target label, $y$. Since the estimator is simple, it will not be very accurate and the prediction will have a certain residual error, $y-F_1(X)$. In the second iteration, we train another weak estimator, $h_1(X)$ not on the desired output, but on residual error of the previous step to create a new, improved, estimator $F_2(X) = F_1(X) + h_1(X)$, since in the ideal case $h_1(X) = y -F_1(X)$ and we would have the desired prediction. However, each step, $i$, will have a residual error $y-F_i(X)$, and so the process can be repeated until the desired accuracy (or maximum number of iterations) is reached. The final GBM output will then be a function of all the steps, $F(X)=F_m(X)= F_{m-1}(X)+h_{m-1}(X)$, where $m$ is the total number of steps. This presentation is schematic and can be generalised to any problem with a differentiable loss function.\footnote{The term \emph{boosting} arises from iteratively boosting the performance using the output of another weak learner. The term \emph{gradient} comes from the observation that fitting successive weak estimators on the residuals of the previous iteration is equivalent to a gradient descent step in a function space spanned by the weak learners.} GBM have been known in the literature to be powerful estimators for tabular data, which we explore in this work. The usual implementation of GBM uses simple trees at each step. However, this requires sorting the data at each iteration, making tree-based GBM computationally heavy for datasets larger than a few tens of thousands examples, which is our case. To mitigate this, a histogram-base variant has been developed that sorts and binarises (i.e., assign each data point to the bins of the histogram) the data once, producing orders of magnitude speed improvements in training and prediction. In this work, we used the histogram-based GBM implemented by \verb|scikit-learn|'s \verb|HistGradientBoostingClassifier| (HGBC).

In a preliminary study, we compared HGBC against other GBM implementations and \verb|scikit-learn|'s RF, observing all GBM to outperform the RF when computing the Area Under the Curve (AUC) of the Receiver Operating Characteristic (ROC) evaluated on the validation set. We then tuned the hyperparameters of the HGBC, observing minimal improvements of its discriminating performance, for which reason we decided to keep the default hyperparamters for the rest of the study presented herein.\footnote{The performance improvements only impacted at most the third significant digit of the validation ROC AUC, which is arguably at the level of statistical fluctiation of the dataset itself, and therefore meaningless.}

\section{Analysis}

In this section, we present the multiclassification analysis on the two generated sets: the first generated with the OGLE-II timestamps, and the second generated with ideal regular daily cadence. The purpose of performing two analyses is to assess whether and how the discrimination power is affected by the cadence.

\subsection{OGLE-II Timestamps}

\begin{figure}
	\centering
	\includegraphics[width=.45\textwidth]{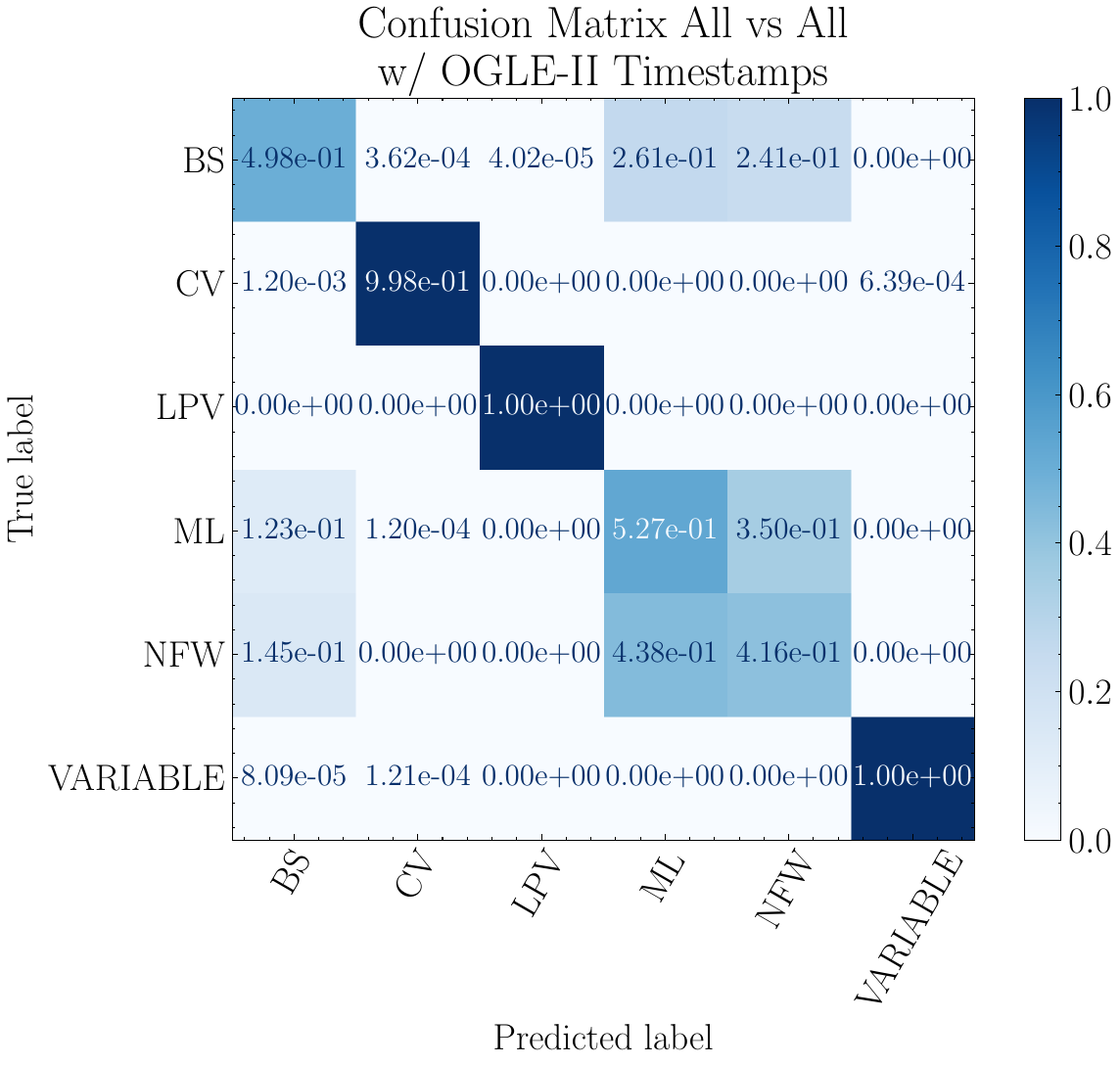}
	\caption{Confusion matrix for the six-way All vs. All multiclassification performed by the HGBC using the dataset generated with OGLE-II timestamps. The entries are rounded to three significant digits.}
	\label{Fig:confusion_matrix_all_vs_all_OGLEII}
\end{figure}

In~\cref{Fig:confusion_matrix_all_vs_all_OGLEII} we present the confusion matrix of the six-way (All vs. All) multiclassification performed by the HGBC. We observe that the non-microlensing events -- CV, LPV, VARIABLE -- have minimal to nonexisting overlap with any class. Conversely, the HGBC prediction for the microlensing events -- ML, BS, and NFW -- has significant overlap. However, we observe that
BS suffer from considerable less contamination from ML and NFW events than these two do of each other. This suggests that BS events are the easiest to isolate, and therefore to detect, of the three microlensing cases.

\begin{figure}
	\centering
	\includegraphics[width=.475\textwidth]{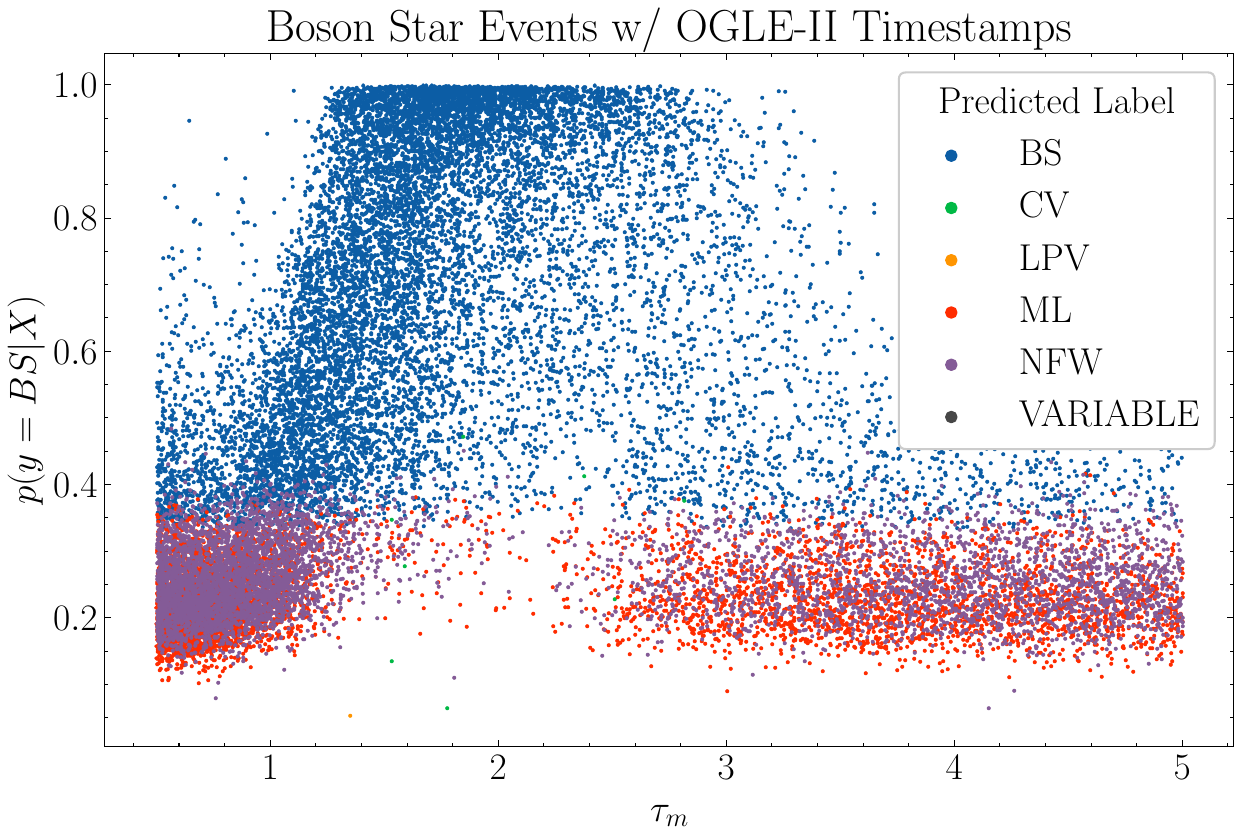}
	\caption{Probability of a BS light curve being correctly identified as one by the HGBC versus the Boson Star $\tau_m$ parameter, using the datset generated with OGLE-II timestamps.}
	\label{Fig:pred_bs_vs_tm}
\end{figure}

Since~\cref{Fig:confusion_matrix_all_vs_all_OGLEII} suggests that BS events are easy to isolate, it is important to understand the nature of the BS we identify. All BS follow the same mass profile, $m(t)$, which can be wider or narrower, depending on the parameter $\tau_m$, associated with the radius of the BS. In~\cref{Fig:pred_bs_vs_tm} we show how the probability of a BS being identified as such by the HGBC depends on $\tau_m$. We find that there is a \emph{sweet spot} at $\tau_m\simeq 2$ to maximise the correct identification of BS, with some some BS light curves being classified with high confidence, i.e. with $P(y=BS|X)\simeq 1$.
	{This also motivates a posterior  to our choice of interval $ 0.8 < \tau_m < 3$ in fig.~\ref{Fig:sensitivity}.}
In~\cref{Fig:10_most_distinctive_BS_lc_OGLEII} we present the 10 BS light curves with the highest value of $P(y=BS|X)$, and we observe that they all exhibit the three peak magnification profile produced by caustics that one expects from BS sources.

\begin{figure}
	\centering
	\includegraphics[width=.4\textwidth]{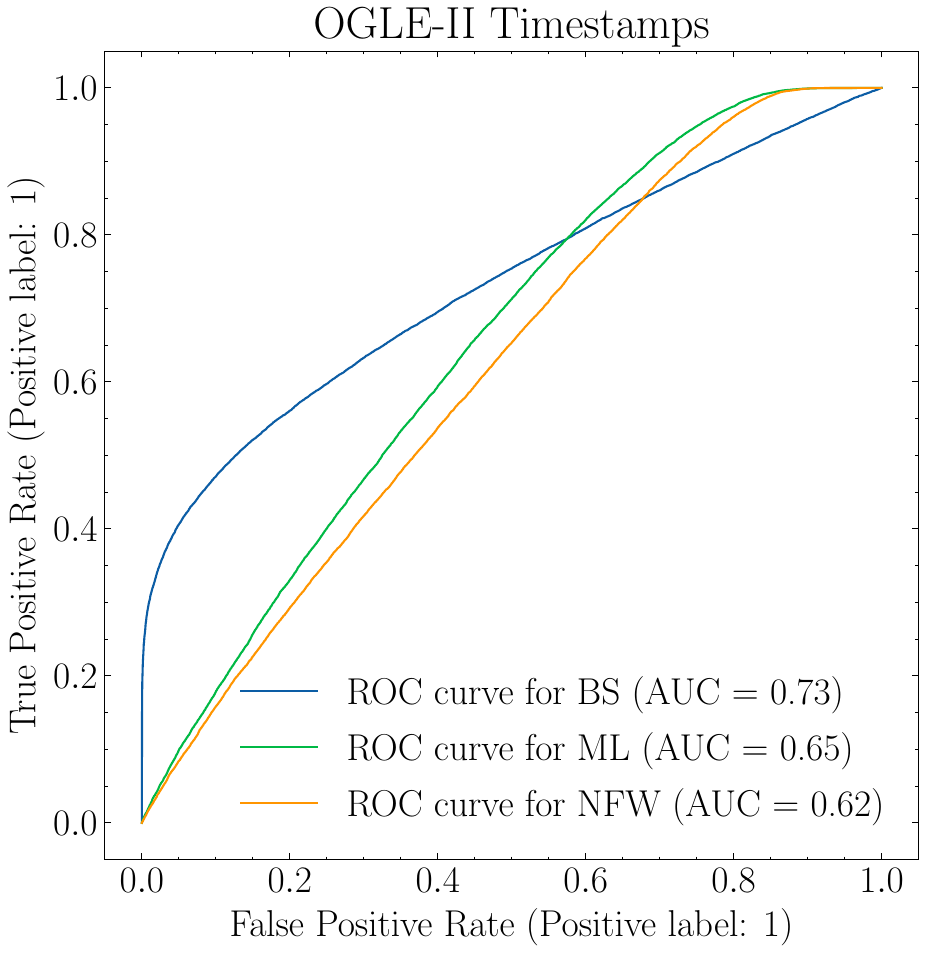}
	\caption{ROC curves and their areas obtained from a HGBC trained on the three-way multiclassification task using the dataset generated with OGLE-II timestamps.}
	\label{Fig:roc_curves_focus}
\end{figure}

Our analysis so far has shown that the discrimination between microlensing sources (ML, BS, NFW) and other sources of light curves (CV, LPV, VARIABLE) is a relatively easy task when OGLE-II cadence is observed, with almost nonexisting overlap between these two broad classes. As such, we now focus on the difficult three-way classification focused on discriminating between the three microlensing sources. In~\cref{Fig:roc_curves_focus} we present the ROC and the AUC of each class (versus the other two) for a HGBC trained on this three-way multiclassification problem.\footnote{The AUC of a ROC curve can then be seen as an average performance over all possible thresholds set on the predicted probability used to predict the class.} The advantage of analysing a ROC curve, and its area, over a classification matrix is that the ROC curve captures the classification performance over all possible output threshold cuts, while the confusion matrix only shows the true and false positives of the assigned predicted class as the one that has maximal probability. For example, in~\cref{Fig:roc_curves_focus} we can see that BS light curves can be isolated with high purity (i.e. with False Positive Rate around $0.0$), while that is not possible for neither ML nor NFW light curves, again reinforcing that BS light curves are easier to identify than the remaining microlensing sources due to their unique three-peak profile. Furthermore,~\cref{Fig:roc_curves_focus} suggests that point-like microlensing light curves are only slightly easier to identify than NFW ones, since their ROC curve has a True Positive Rate greater than that of the NFW for all values of the False Positive Rates. This is also evidenced by the ML AUC, $0.65$, which is greater than that of the NFW, $0.62$.

\begin{figure}
	\centering
	\includegraphics[width=.49\textwidth]{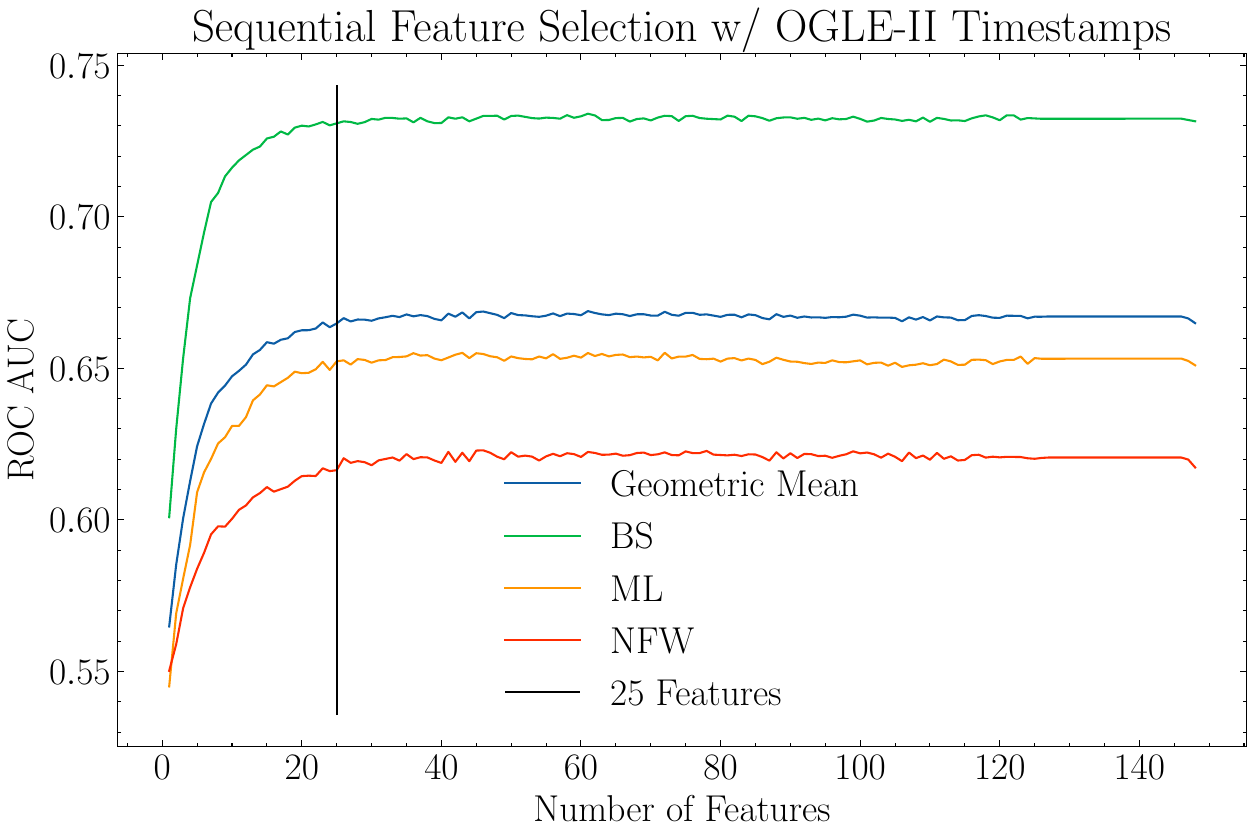}
	\caption{Backward SFS (features reduced from right to left) for a HGBC trained on the three-way multiclassification task using the dataset generated with OGLE-II timestamps.}
	\label{Fig:roc_auc_vs_number_of_features}
\end{figure}

Having been able to isolate the BS light curves from other microlensing sources, we now turn to interpreting our results. Unfortunately, although HBGC are incredibly powerful estimators they do no provide a clear way of interpreting their prediction, a common challenge when using machine learning in highly dimensional multivariate analyses. In this study, we implemented a backward Sequential Feature Selection (SFS) loop to assess which of the 148 features are relevant for this classification task. The backward SFS loop starts by fitting an HGBC using all 148 features and evaluating its performance on the validation set. Then we train 148 HGBC on all the possible 148 subsets {comprised} of 147 features (i.e., with one feature removed) and evaluate their performance on the validation set.  Keeping only the best set of 147 features, the same step can be done for the 146 subsets and so one, for a total of 148 iterations totaling 11026 HGBC trained on 11026 subset of features, which is a feasible study given the increased training speed offered by the HGBC.\footnote{Noting that a set of 148 elements has $2^{148}-1$ non-empty subsets, 11026 is far fewer than the total possible combinations of features.} In~\cref{Fig:roc_auc_vs_number_of_features} we show the performance of the HGBC on the three-way microlensing  focused multiclassification for each of the possible sources and their geometric mean. We see that only around 25 features derived from the light curve time series and its derivatives, out of the 148, are relevant for this task.\footnote{It is important to point out that while this methodology is aimed at removing uninformative features, it will also remove redundant, i.e. highly correlated, features. Therefore, the final set of features is only unique up to fluctuations of the data and the stochastic initialisation of the HGBC. To mitigate this, we fixed the seed of the run, but modifications of the training and validation set could still hold a different set of features.} We also note that the discrimination performance degrades for all the three classes at the same stage, suggesting that the final 25 features are relevant for all three classes. The list of the final 25 features by survival ranking is presented in~\cref{tab:top-25-features-backward-sfs}.

\begin{figure}
	\centering
	\includegraphics[width=.475\textwidth]{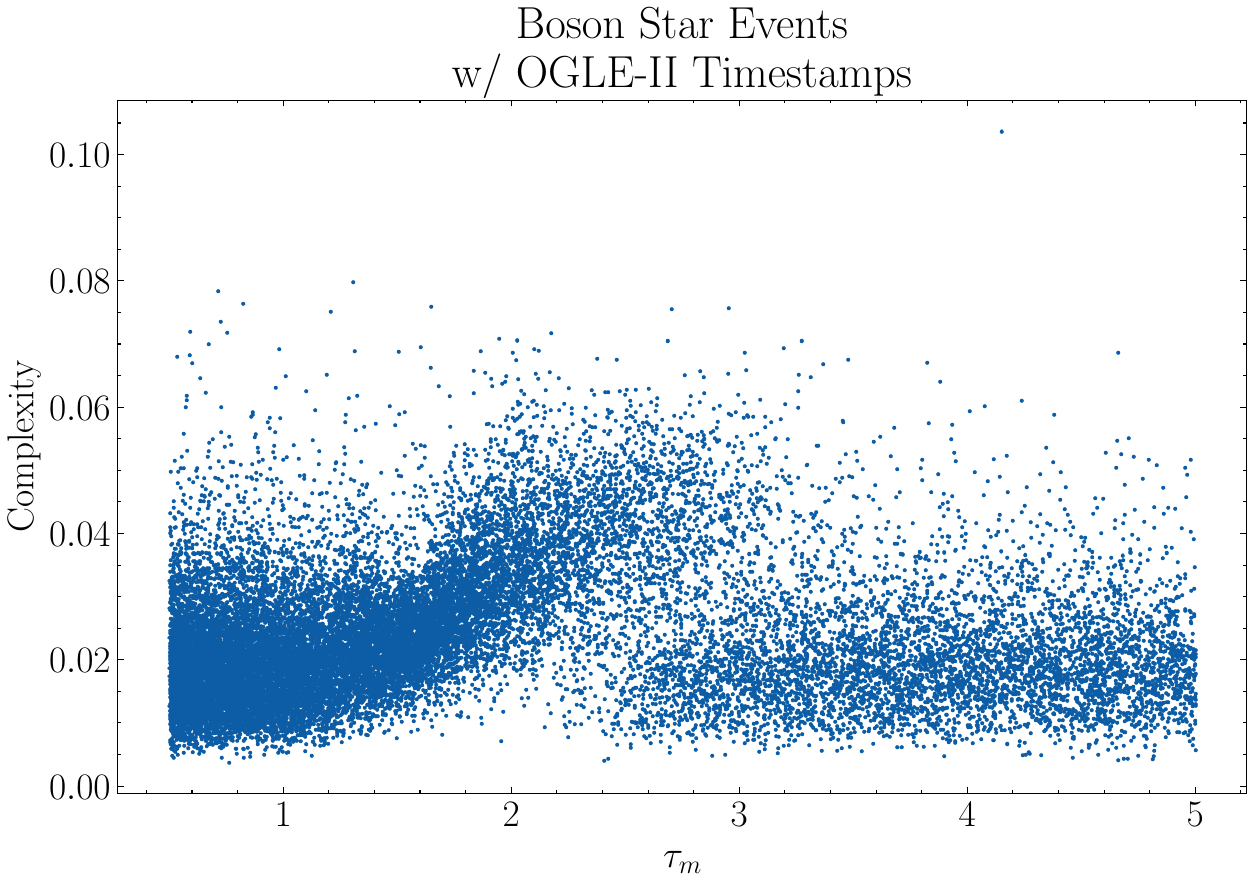}
	\caption{Complexity vs $\tau_m$ scatter for BS events with OGLE-II timestamps.}
	\label{Fig:complexity_vs_t_m_BS}
\end{figure}

For the dataset using OGLE-II timestamps, which is the subject of the analysis in this section, we find that the \emph{complexity} of the time series is the most important feature to separate the light curves from the three microlensing sources, surviving the whole backward selection loop until there is only one. In~\cref{Fig:complexity_vs_t_m_BS} we show the distribution of the time series complexity for BS against the $\tau_m$, where we witness a boost in the time series complexity for $\tau_m\simeq 2$, the same sweet spot identified above in~\cref{Fig:pred_bs_vs_tm}, suggesting that the complexity of the time series of a BS light curve is a driving feature for it to be correctly identified as a BS. The same is not observed for the NFW events, which we do not show for the sake of presentation tidiness.

In~\cref{Fig:10_most_distinctive_BS_lc_OGLEII} we show the 10 most distinctive BS light curves, i.e. the 10 BS light curves with highest $P(y=BS|X)$ as identified by the HGBC trained on the three-way multiclassification task on the OGLE-II dataset. These light curves all exhibit very clear three magnitude peaks, a hallmark feature arising from the caustics produced by the extended nature of the lens.

\subsection{Regular Daily Cadence Timestamps}

So far our analysis has focused on light curves simulated using OGLE-II timestamps. This reflects well the sensitivity to extended objects in a realistic microlensing survey, with its given observational constraints. However, one
might wonder to what extent the conclusions previously drawn are sensitive to the observation cadence details, especially its irregularity. To address this, we conduct the analysis with light curves simulated with regular daily cadence, i.e. where observations are taken exactly 24 hours apart.

\begin{figure}
	\centering
	\includegraphics[width=.475\textwidth]{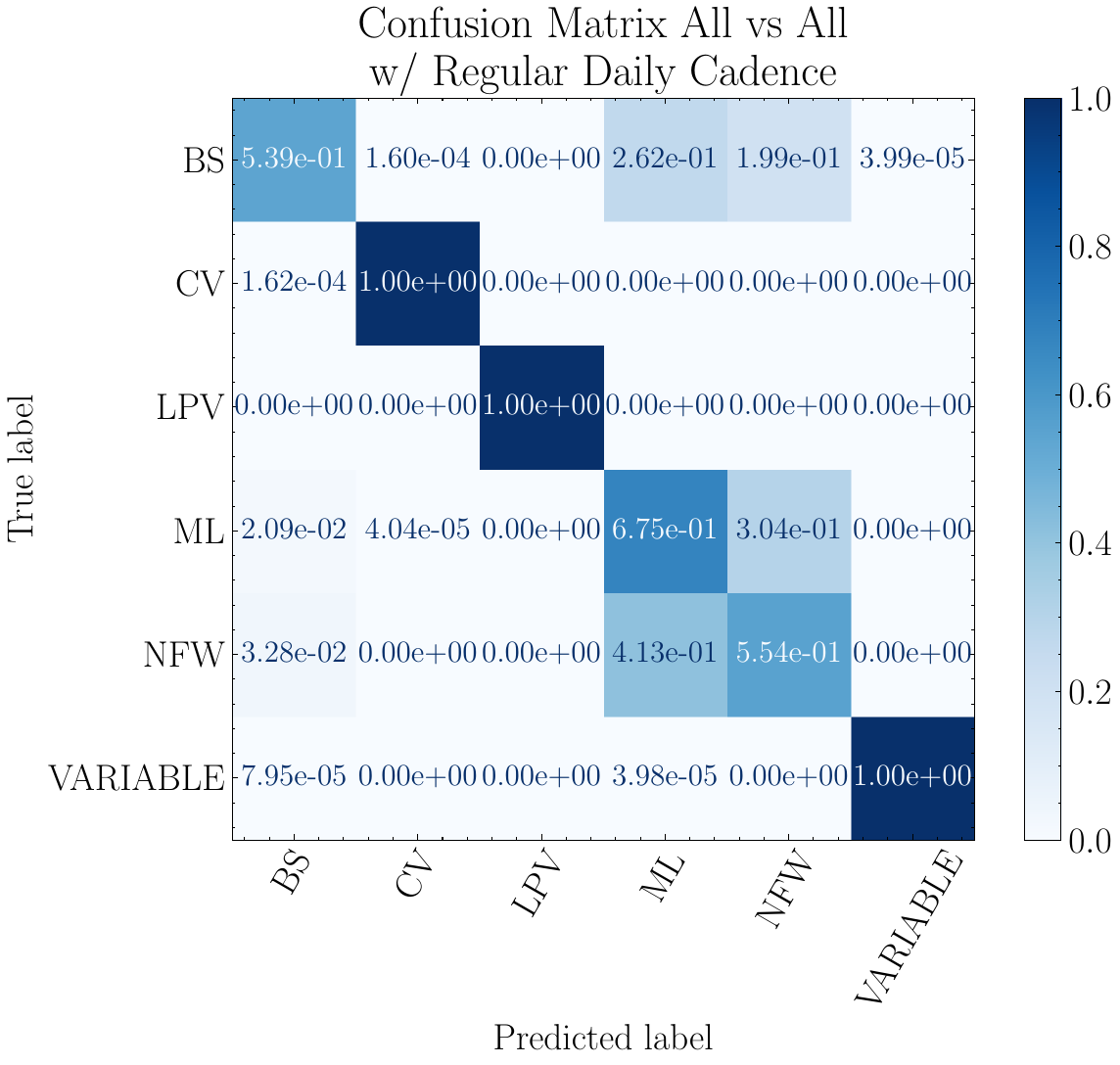}
	\caption{Confusion matrix for the six-way All vs. All multiclassification performed by the HGBC using the datset generated with regular daily cadence timestamps. The entries are rounded to three significant digits.}
	\label{Fig:confusion_matrix_all_vs_all_regular-cadence}
\end{figure}

We begin with the six-way All vs All multiclassification task. In~\cref{Fig:confusion_matrix_all_vs_all_regular-cadence} we show the confusion matrix obtained using the HGBC. Although similar to its counterpart with OGLE-II timestamps,~\cref{Fig:confusion_matrix_all_vs_all_OGLEII}, it has some noticeable differences. First, we notice that the contamination in BS positive predictions has decreased from $\mathcal{O}(10\%)$ down to $\mathcal{O}(1\%)$, suggesting an improved capacity of the HGBC in producing a pure sample of BS light curves. Secondly, we notice that the cross misclassification between NFW and ML light curves has also decreased, improving upon the OGLE-II timestamps case.

\begin{figure}
	\centering
	\includegraphics[width=.475\textwidth]{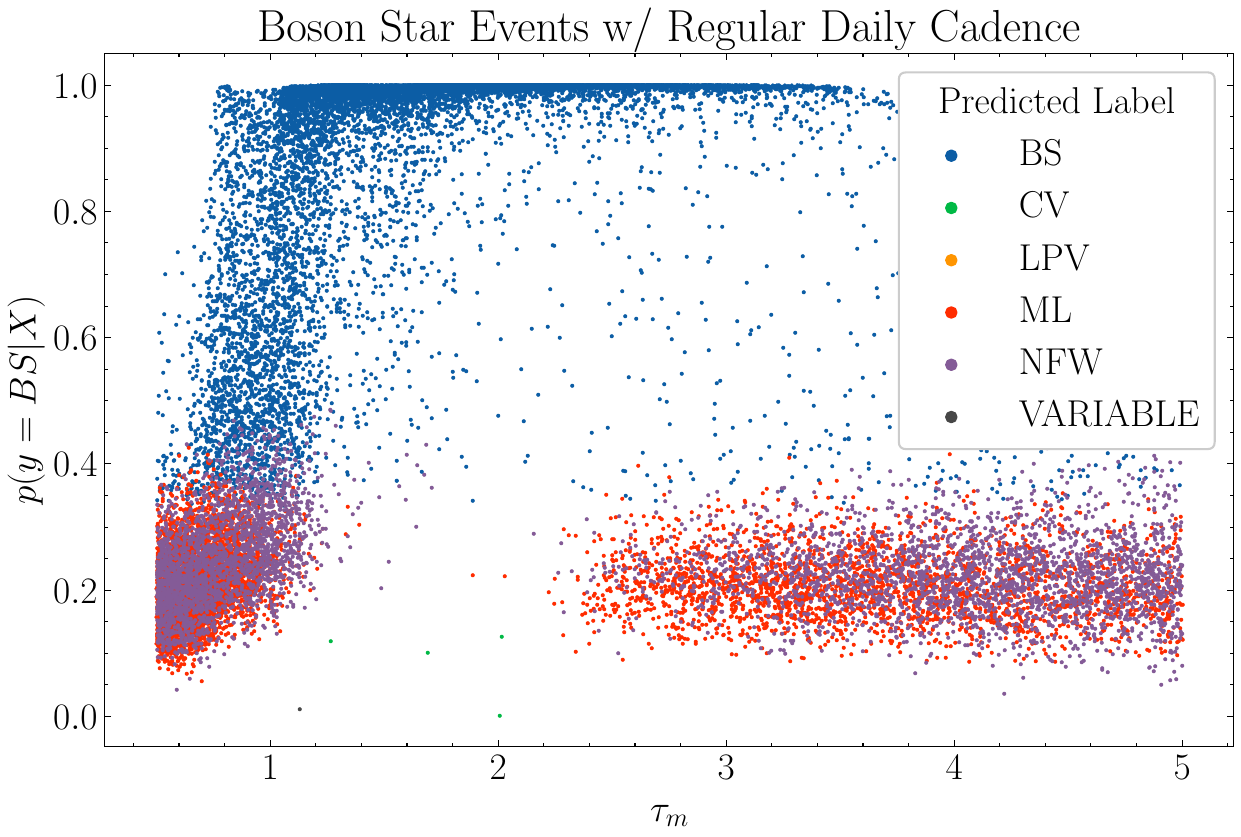}
	\caption{Probability of a BS light curve being correctly identified as one by the HGBC versus the Boson Star $\tau_m$ parameter, using the datset generated with regular cadence timestamps.}
	\label{Fig:pred_BS_vs_t_m_for_BS_light curves_regular-cadence}
\end{figure}

In~\cref{Fig:pred_BS_vs_t_m_for_BS_light curves_regular-cadence} we revisit the probability of a BS light curve being classified as such versus the BS $\tau_m$, but this time for the regular cadence case. We see that the previously identified~\emph{sweet spot} at $\tau_m\simeq 2$ has now almost no missclassifications into other classes, showing how a regular cadence can lead to an even better identification of BS light curves; this is due to a better resolution of the light curve being able to capture the three magnification peaks.

\begin{figure}
	\centering
	\includegraphics[width=.4\textwidth]{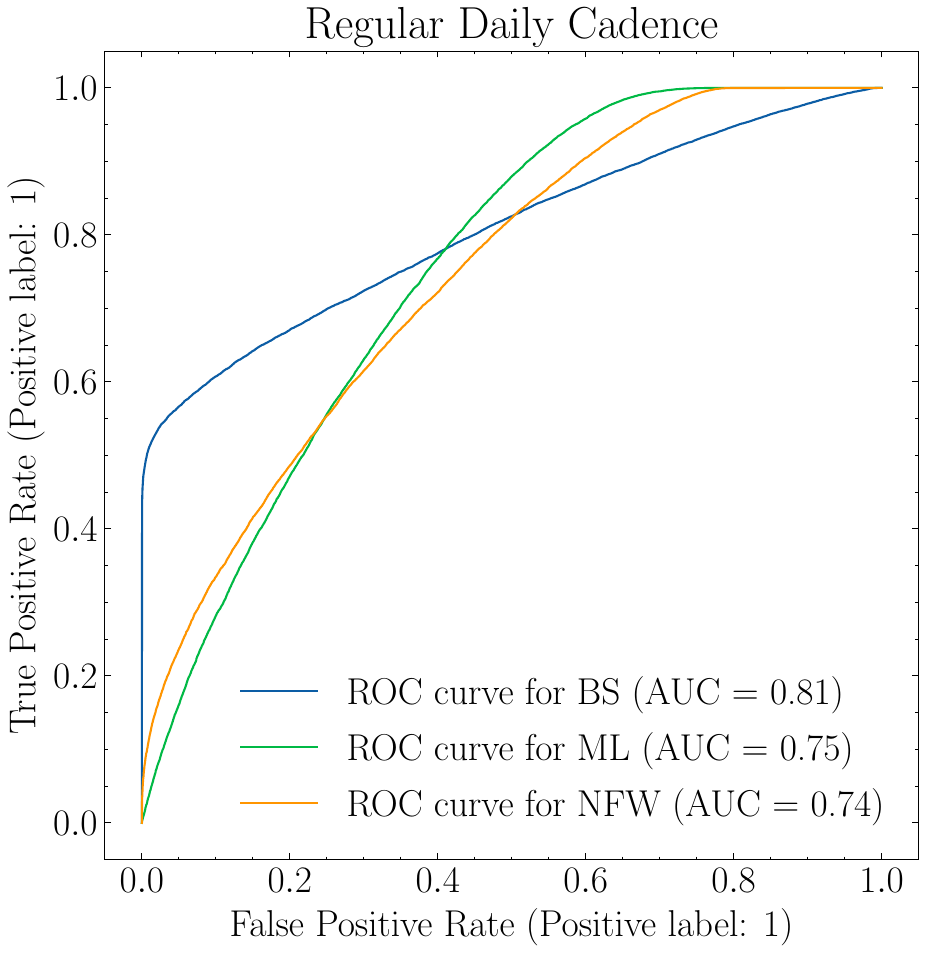}
	\caption{ROC curves and their areas obtained from a HGBC trained on the three-way multiclassification task using the datset generated with regular daily cadence timestamps.}
	\label{Fig:roc_curves_focus_regular-cadence}
\end{figure}

As before, we now focus on the three-way multiclassification problem focused on the three microlensing sources. The ROC curves for each of the three classes and their respective AUC can be seen in~\cref{Fig:roc_curves_focus_regular-cadence}. Again, we see that there are some noticeable differences to the OGLE-II case shown in~\cref{Fig:roc_curves_focus}. The first thing to notice is that all AUC have increased compared to the OGLE-II timestamps case, indicating an easier discrimination when using regular daily cadence timestamps. Next, we see how the ROC curve for the BS is significantly more peaked for small False Positive Rate, providing further evidence that the BS light curves are better classified with regular daily cadence. More interestingly, however, is how the ML and NFW ROC curves no longer follow the same trend as in the OGLE-II case. More precisely, we can observe that they cross, as before they did not. This happens halfway through the curve, with the NFW ROC curve having the \emph{upper hand} over the ML ROC curve for lower values of the False Positive Rate, with an added feature that the ROC curve for the NFW can have non-vanishing True Positive Rate at low False Positive Rate values. This implies that with regular daily cadence it is possible to isolate NFW light curves with little to no contamination of any other class, something that was impossible to achieve using the OGLE-II cadence.

\begin{figure}
	\centering
	\includegraphics[width=.475\textwidth]{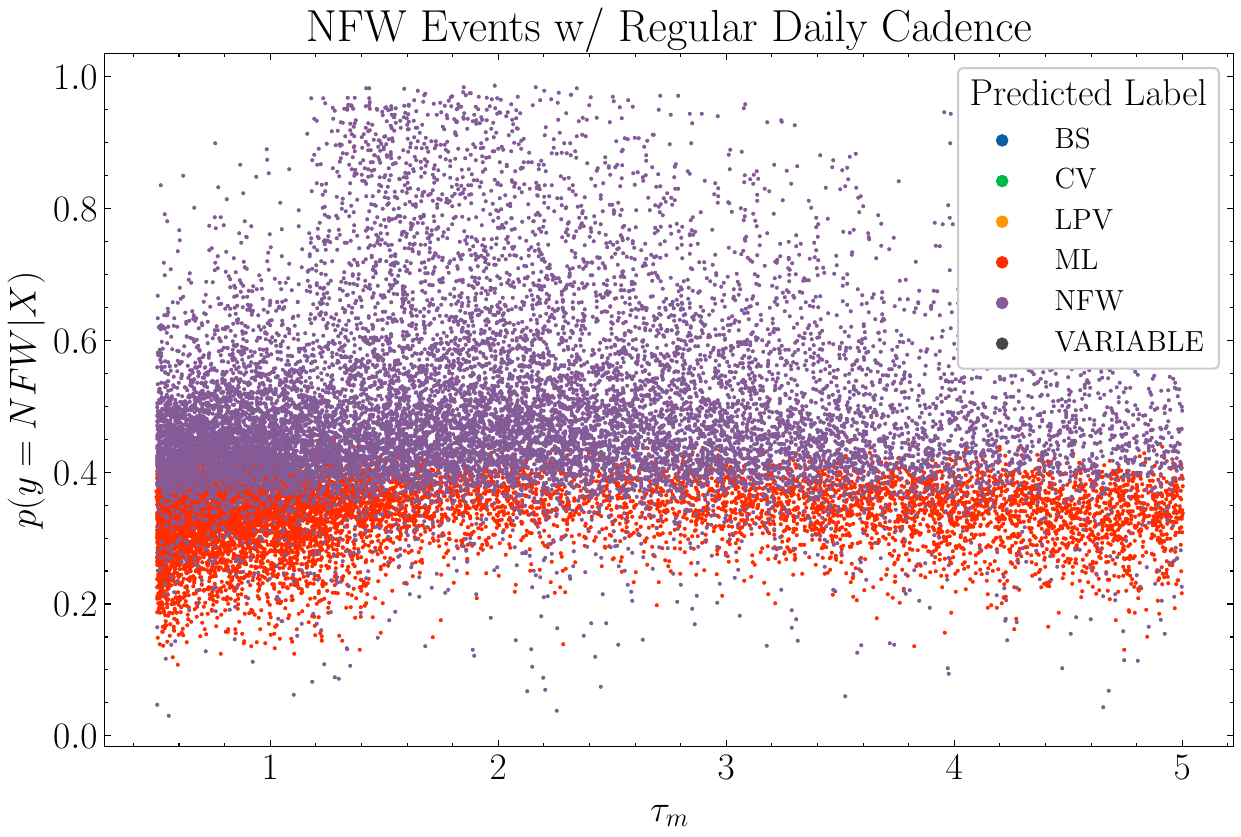}
	\caption{Probability of a NFW light curve being correctly identified as one by the HGBC versus the NFW $\tau_m$ parameter, using the dataset generated with regular daily cadence timestamps.}
	\label{Fig:pred_NFW_vs_t_m_for_NFW_light curves_regular-cadence}
\end{figure}

The previous result points at the possibility {in principle} of completely isolating NFW light curves. It is then important to understand the nature of the NFW light curves that we can isolate. In~\cref{Fig:pred_NFW_vs_t_m_for_NFW_light curves_regular-cadence} we show how the probability for an NFW light curve to be correctly identified as such can vary with the NFW $\tau_m$ parameter. Although less noticeable than what we saw before for the BS light curves, we can observe a \emph{sweet spot} for correct classification at $1 \lesssim \tau_m \lesssim 2$, with some light curves being assigned $P(y=NFW|X)\simeq 1$. We notice that the equivalent scatter for the OGLE-II timestamps, not shown here to declutter the presentation, does not exhibit this pattern, suggesting that there is important information in the light curve that can only be obtained with regular cadence.

\begin{figure}
	\centering
	\includegraphics[width=.475\textwidth]{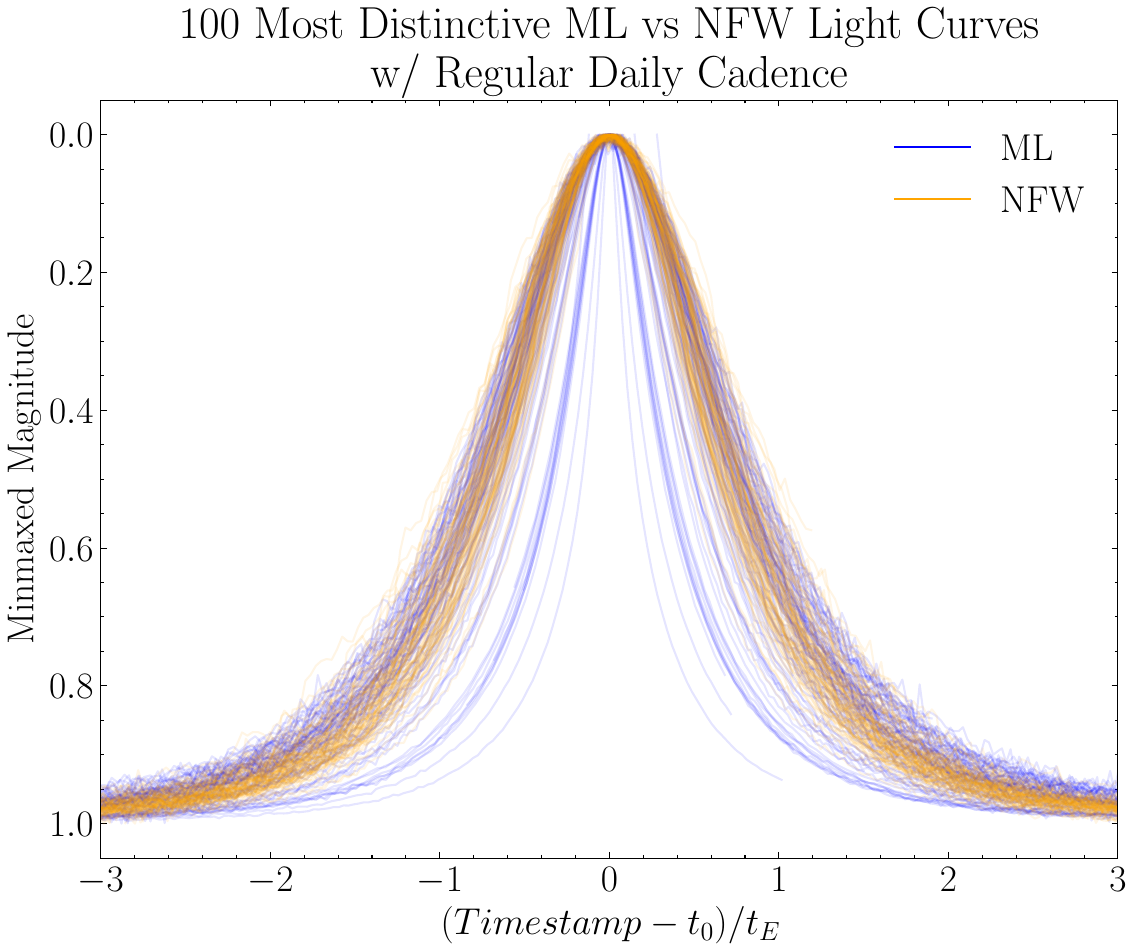}
	\caption{Comparison between the 100 most easily identifiable ML curves and the 100 most easily identifiable NFW curves.
	}
	\label{Fig:100_most_distinctive_ML_vs_NFW_lc_regular-cadence}
\end{figure}

Contrary to BS light curves, NFW light curves do not have a clear visual profile compared to point-like microlensing sources. In~\cref{Fig:100_most_distinctive_ML_vs_NFW_lc_regular-cadence} we compare the 100 most easily identifiable point-like microlensing curves, i.e. those with the highest $P(y=ML|X)$, against the 100 most easily identifiable NFW curves, i.e. those with the highest $P(y=NFW|X)$, where the probabilities are obtained from the HGBC trained on the three-way multiclass classification on the regular cadence timestamps dataset. We normalised the magnitudes by min-maxing their values to fall under the $[0,1]$ interval to aid visual comparison. Although the difference is very nuanced, the NFW curves tend to be \emph{narrower} than the ML curves. This subtlety explains why regular cadence is so important for identifying NFW light curves, as one needs a better \emph{resolution} of the light curve profile to be able to identify this nuance, which proved impossible when using the OGLE-II timestamps.

\begin{figure}
	\centering
	\includegraphics[width=.49\textwidth]{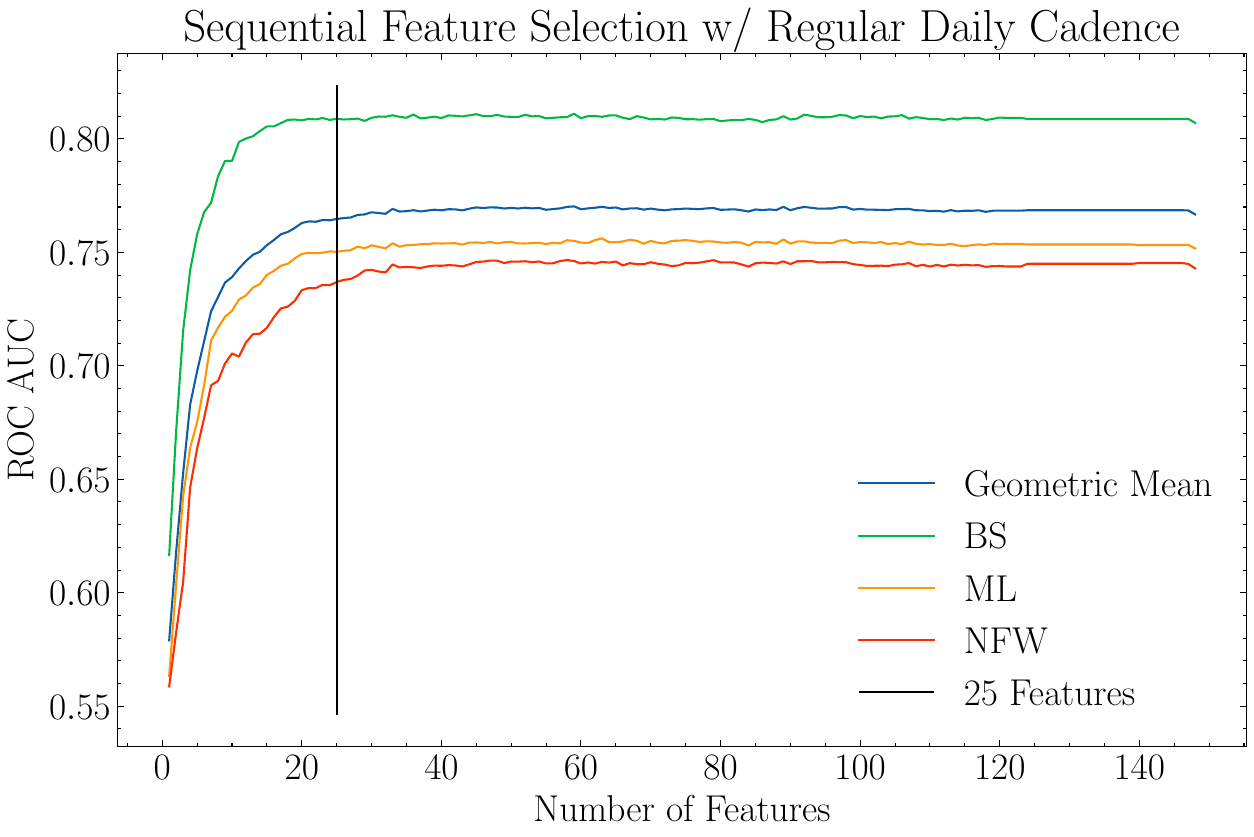}
	\caption{Backward SFS (features reduced from right to left) for a HGBC trained on the three-way multiclassification task using the dataset generated with regular daily cadence timestamps.}
	\label{Fig:roc_auc_vs_number_of_features_regular-cadence}
\end{figure}

Finally, we perform a backward SFS for the regular daily cadence dataset using a three-way HGBC focused on the microlensing classes. The discrimination performance versus the number of features is shown in~\cref{Fig:roc_auc_vs_number_of_features_regular-cadence}, where we observe a similar trend than that already discussed for the OGLE-II timestamps dataset with the ROC AUC degrading significantly below 25 features. However, it is worth noting that the performance degrades first for the NFW and ML cases, with the ROC AUC associated with the classification of the BS light curves staying at its maximum value until around 15. This suggests that there are less relevant features to distinguish BS light curves from the other classes than there are to correctly identify the others. In~\cref{tab:top-25-features-backward-sfs} we present the top 25 features, and we observe considerable overlap with the OGLE-II study. Perhaps curiously, we count 10 of the features to have been computed on the derivative of the time series, whereas for the OGLE-II the number is six, possibly hinting at the importance of the derivative of the time series in identifying the nuanced differences in shape between the ML and the NFW light curve magnification peak.

\section{Discussion}
In this work we have studied the observation of microlensing signatures due to extended dark objects in time series data, and its distinction from other signals, most importantly point-like lenses. We have focused on two lens profiles, each exemplifying a class of objects: the profile of an NFW-subhalo is more peaked, whereas boson stars have a more diffuse profile. As expected, the boson stars can be confidently distinguished from point-like lenses, for $0.8 \lesssim \tau_m (=r_{\rm lens}/r_E) \lesssim 3$. This is owed to the characteristic caustics in the light curves for these objects.

As an exercise, we studied how the regularity of the time series cadence affects the confidence of the detections.
Interestingly, for regular daily cadence we also find confident detections of NFW-subhalos. These detections occur for $ 0.9 \lesssim \tau_m \lesssim 4$, despite the fact that no caustics are observed in the light curves. Though the regularity of (daily) observations is dependent on many conditions that are beyond the observer's control, this is an interesting observation which warrants investigation for other microlensing scenarios and targets.

Stellar binaries or exoplanets orbiting a lensing body may give rise to perturbations and caustics in the light curve. Unlike for extended dark objects, these are asymmetric or one-sided features (see e.g. \cite{khakpash2021classifying}). Microlensing has allowed for the discovery of over 100 exoplanets, particularly of near-terrestial size \cite{ExoplanetArchive}. In this work, we have not considered these signatures, which may give rise to confusion with boson star identifications with low significance. We will leave such an analysis for future work.

We performed the analysis in this work by extending the MicroLIA algorithm, which utilises 148 features derived from the light curves in a single band to distinguish between objects. Our SFS analysis showed that of these, only 25 were needed for the optimal classification. For our regular cadence analysis, further features could help distinguish between NFW-subhalos and point-like lenses.
In future work, we will study the potential benefits of learning directly on the light curves.

	{We note that in our analysis we have assumed a Gaussian noise model, which may not be realistic.
		In future work towards the application of our methodology, further noise models should be considered. These could lead to larger misclassification between classes of events.}

Microlensing surveys typically only release data on candidate events, which have passed through a strict selection in which our extended dark object light curves were likely cut. Exceptions include the UKIRT Microlensing Survey \cite{UKIRT,shvartzvald2017ukirt} and the VISTA Variables in the Via Lactea Survey \cite{minniti2010vista}, which we plan to analyse in future work.
Upcoming microlensing opportunities include the launch of infrared astronomy experiments in the mid-2020s.
In particular, the Nancy Grace Roman Space Telescope (previously WFIRST; a space mission by NASA) has as a key objective to discover exoplanets through the microlensing technique \cite{Spergel:2015sza}, but will also probe primordial black holes \cite{Fardeen:2023euf} and extended dark objects \cite{DeRocco:2023hij}.

Finally, for some surveys, the finite extend of the source becomes important.
This is the case, for example, for the Subaru-\acro{hsc} survey of M31~\cite{Subaru}: because of its sensitivity to small transit times (and hence small Einstein radii), the
angular extent of source stars corresponds to a distance at the lens larger than the Einstein radius, suppressing the magnification relative to point-like sources \cite{WittMao,Montero-Camacho:2019jte,SantaCruzFiniteSource,Croon:2018ybs}. We leave an analysis of the magnification curves with finite source effects for future work.

\section*{Software}

For the computation of the mass profile of the extended sources and the sensitivity estimation, we used \verb | Mathematica | version 12.

The dataset generation, machine learning training, and the final analysis were performed in \verb|python 3.9.18|, making use of several packages, of which we make especial note:
\begin{itemize}
	\item Light curve simulation and time series feature extraction was performed using MicroLIA. We used the code directly from the MicroLIA github repository, as it includes considerable changes to the code provided with the packaged version \verb|2.6.0|. More precisely, the version used makes proper use of the derivative time series, as well as its errors as statistical weights. These two fixes were contributions of this work.
	\item \verb|scikit-learn 1.3.2| for HGBC implementation.
	\item \verb|mlxtend 0.23.0| for the backwards SFS.
	\item \verb|SciencePlots 2.1.1| for plotting.
\end{itemize}

The full list of packages and their versions can be found in the \verb|requirements.txt| file included in the code repository hosting the code used in this work.\footnote{\url{https://gitlab.com/miguel.romao/microlensing-extended-objects-machine-learning}}

\section*{Acknowledgements}
We thank Daniel Godines help regarding MicroLIA and feedback on the manuscript, and Steve Abel and Nirmal Raj for useful discussions.~MCR and DC are supported by the STFC under Grant No.~ST/T001011/1.

\appendix

\section{Backward Sequential Feature Selection Results}

In~\cref{tab:top-25-features-backward-sfs} we collect the top 25 features obtained by the backward Sequential Feature Selection (SFS), for both the OGLE-II and the regular daily cadence timestamps datasets. The details of the features can be seen in~\cite{godines2019machine} and in MicrioLIA's code. The feature rank is assocaited with how long it survived the SFS loop. E.g., a rank $1$ means that the feature is present in all subsets down to the last subset of a single feature, whereas a rank $n$ means that the feature is present up until the subset of size $n$, but is then removed and is not present in smaller subsets. This set of features is only unique up to fluctuations of the data and the stochastic variability associated to HGBC training. In principle, any feature presented in this table could be replaced by a highly correlated counterpart, and as such the list presented herein should be interpreted as the maximal set of mutually uninformative features obtained using our training and validation sets.
\begin{table*}[t]
	\centering
	\begin{tabular}{c|c | c}
		\hline\hline
		Feature Rank & OGLE-II                           & Daily Regular Cadence                \\\hline
		1            & complexity                        & stetsonJ (derivative)                \\
		2            & median buffer range               & half mag amplitude ratio             \\
		3            & ratio recurring points            & FluxPercentileRatioMid80             \\
		4            & amplitude                         & median buffer range                  \\
		5            & medianAbsDev                      & count below                          \\
		6            & FluxPercentileRatioMid80          & medianAbsDev                         \\
		7            & count below                       & check max last loc                   \\
		8            & cusum                             & longest strike below                 \\
		9            & shapiro wilk                      & complexity                           \\
		10           & mean change                       & PercentDifferenceFluxPercentile      \\
		11           & longest strike above (derivative) & longest strike above                 \\
		12           & mean second derivative            & shapiro wilk (derivative)            \\
		13           & number of crossings               & FluxPercentileRatioMid65             \\
		14           & LinearTrend                       & number of crossings                  \\
		15           & longest strike below (derivative) & number cwt peaks                     \\
		16           & sample entropy                    & number cwt peaks (derivative)        \\
		17           & FluxPercentileRatioMid65          & time reversal asymmetry (derivative) \\
		18           & mean change(derivative)           & FluxPercentileRatioMid35             \\
		19           & longest strike above              & peak detection(derivative)           \\
		20           & mean n abs max (derivative)       & mean n abs max (derivative)          \\
		21           & FluxPercentileRatioMid20          & longest strike above (derivative)    \\
		22           & quantile (derivative)             & check min last loc                   \\
		23           & check min last loc                & quantile (derivative)                \\
		24           & Gskew                             & sample entropy (derivative)          \\
		25           & check max last loc (derivative)   & below3 (derivative)                  \\
		\hline\hline
	\end{tabular}
	\caption{Top 25 features, ranked by \emph{survivability} of the backward Sequential Feature Selection loop. Features computed in the derivative of the time series are identified as such. See~\cite{godines2019machine} for details about the features.}
	\label{tab:top-25-features-backward-sfs}
\end{table*}

\bibliography{main}

\end{document}